\documentclass[11pt,letterpaper]{article}
\usepackage[letterpaper,hmargin=1in,vmargin=1.25in]{geometry}
\usepackage{amsthm,amsmath,amscd,
amssymb,mathrsfs,ccaption,verbatim,graphicx,color,layout}

\def\Gen{\operatorname{Gen}}
\def\params{\operatorname{\frak p}}

\def\PI{\operatorname{PI}}
\def\AD{\operatorname{AD}}

\def\Adv{\operatorname{\cal A}}
\def\Ch{\operatorname{Ch}}
\def\ADVANTAGE{\operatorname{Adv}}

\def\bs{\backslash}
\def\Lam{\Lambda}
\def\PIE{\operatorname{PIE}}
\def\PIR{\operatorname{PIR}}
\def\PIC{\operatorname{PIC}}
\def\E{\operatorname{E}}

\def\CMR{\operatorname{\it CMR}}
\def\FCMR{\operatorname{\it FCMR}}
\def\FNCMR{\operatorname{\it FNCMR}}

\def\diff{\operatorname{diff}}

\def\BioPerf{\operatorname{BP}}
\def\ToPerf{\operatorname{TP}}
\def\Div{\operatorname{Div}}
\def\AL{\operatorname{\AL}}
\def\PAL{\operatorname{\PAL}}
\def\Samp{\operatorname{Samp}}

\def\FAR{\operatorname{\it FAR}}
\def\rMR{\operatorname{\it rMR}}
\def\MR{\operatorname{\it MR}}
\def\FMR{\operatorname{\it FMR}}
\def\FNMR{\operatorname{\it FNMR}}

\def\match{\operatorname{\it match}}
\def\nonmatch{\operatorname{\it non-match}}

\def\diff{\operatorname{diff}}

\def\sim{\text{{\it sim}}}

\def\B{{\cal B}}

\def\U{{\cal U}}

\def\M{{\cal M}}

\def\vep{\varepsilon}

\def\disp{\displaystyle}

\def\i{\text{\rm i}}
\def\ii{\text{\rm i\hspace{-.2ex}i}}
\def\iii{\text{\rm i\hspace{-.2ex}i\hspace{-.2ex}i}}
\def\randomchosen{\overset{\$}{\leftarrow}}

\def\III{\text{\rm I\hspace{-.2ex}I\hspace{-.2ex}I}}
\def\randomchosen{\overset{\$}{\leftarrow}}
\makeatletter
\def\lapprox{\mathrel{\mathpalette\gl@align<}}
\def\gapprox{\mathrel{\mathpalette\gl@align>}}
\def\gl@align#1#2{\lower.6ex\vbox{\baselineskip\z@skip\lineskip\z@
\ialign{$\m@th#1\hfil##\hfil$\crcr#2\crcr\sim\crcr}}}
\makeatother
\newtheorem{theorem}{Theorem}

\newtheorem{definition}{Definition}

\begin{document}
\markboth{\LaTeXe{} Class for Lecture Notes in Computer
Science}{\LaTeXe{} Class for Lecture Notes in Computer Science}
\title{Relations among Security Metrics\\ for Template Protection Algorithms}
\author{Manabu Inuma
\thanks{National Institute of Advanced Industrial
Science and Technology (AIST), 
1-1-1 Umezono, Tsukuba-shi,
Ibaraki, 305-8568 JAPAN} \and Akira Otsuka $^*$}
\maketitle


\begin{abstract}
Many biometric template protection algorithms have been proposed mainly in two approaches: biometric feature transformation and biometric cryptosystem. Security evaluation of the proposed algorithms are often conducted in various inconsistent manner. Thus, it is strongly demanded to establish the common evaluation metrics for easier comparison among many algorithms. Simoens et al.\cite{SYZBBNP2012} and Nagar et al.\cite{Nagar2010}  proposed good metrics covering nearly all aspect of requirements expected for biometric template protection algorithms. One drawback of the two papers is that they are biased to experimental evaluation of security of biometric template protection algorithms. Therefore, it was still difficult mainly for algorithms in biometric cryptosystem to prove their security according to the proposed metrics. This paper will give a formal definitions for security metrics proposed by Simoens et al.\cite{SYZBBNP2012} and Nagar et al.\cite{Nagar2010} so that it can be used for the evaluation of both of the two approaches. Further, this paper will discuss the relations among several notions of security metrics. 

\end{abstract}
\section{Introduction}
One of the main issues in biometric authentication systems is to protect a biometric template database from compromise. Biometric information is so unique to each user and unchangeable during his or her lifetime. Once biometric template is leaked together with his or her identity, the person will face a severe risk of identity theft. Widely-used template protection systems for biometric authentication systems are tamper-proof hardware-based systems, where biometric template is stored in an ordinary storage as an encrypted form and decrypted only within a tamper-proof hardware  when matching is required. In these systems, even if the database is compromised, biometric information never made public. However,  the drawback of this conventional approach was the requirement of tamper-proof hardware, as it increases the deployment cost especially in high volume matching is required. To overcome this drawback, software-based template protection techniques are proposed recently in many literature\cite{}. Software-based template protection schemes are categorized into 2 approaches\cite{Nagar:2010tg}, feature transformation approach and biometric cryptosystems. Both of them introduces a user-specific key to transform a biometric template into a protected template. 

\subsection{Feature Transformation approach}
Feature transformation approach is first proposed in a paper written by Ratha, Connel and Bolle\cite{Ratha:2001gu} as {\it Cancelable biometrics}. In feature transformation approach, a randomness or key is introduced as a transformation parameter, and each original biometric feature is transformed into a deformed biometric feature. Main advantage of this approach is that it can take benefits from utilizing well-studied high performance algorithms. Thus, the challenge in this approach is to design a transformation function satisfies both (1) that closeness in original biometric feature space should preserve in the transformed feature space and (2) that it is hard to recover the original biometric feature from the transformed feature. On the contrary that feature transformation approach can enjoy the benefit of high-performance algorithms, schemes in this approach tends to have difficulties in theoretical analysis of protection performance such as irreversibility and unlinkability discussed later. Thus, many papers give experimental evidence for security analysis.

Ratha et al. \cite{Ratha:2001gu} introduced the notion of {\it Cancelable biometrics} and proposed several schemes for fingerprint template protection\cite{Ratha:2007it}. Their approach is to displace fingerprint minutiae at different locations according to a irreversible locally smooth transformation. That is, a small change in a minutiae position before transformation leads to a small change in the minutiae position after transformation, but small correlation in minutia positions before and after transformation. Ratha et al.\cite{Ratha:2007it}  evaluated {\it Accuracy} (Section \ref{sect:Accuracy}) for the recognition performance and {\it Irreversibility}(Section \ref{sect:Irr}) for their schemes. They roughly estimated the complexity of irreversibility by the length of its binary representation.

Teoh et al.'s BioHash\cite{Jin:2004kc} and its subsequent papers\cite{Connie:2005wg,Teoh:2006kr,Teoh:2007dd} proposed distance-preserving transformations for biometric feature vectors multiplied with an randomized orthogonal transformation matrix. The randomized orthogonal matrix woks as a user-specific key, it introduces a low false accept rate. Irreversiblity of BioHash is analyzed in \cite{Jin:2004kc} and \cite{Kuan:2005ut}. In \cite{Jin:2004kc}, irreversibility is discussed based on evidences from {\it recognition performance} (Section \ref{sect:RecPerf}) metrics such as {\it accuracy} (Section \ref{sect:Accuracy}), {\it biometric performance} (Section \ref{sect:BioPerf}) and {\it diversity} (Section \ref{sect:Div}). As argued later, for example, in the real world, a fingerprint left on a glass may be abused by a malicious user, then {\it Diveristy} seems to give the complexity of an adversary to find the correct key. However, this discussion only covers a weak adversary whose attacking strategy is specific. A stronger adversary may take other strategies such as finding the correct key by directly inverting the transformation function utilizing the stolen fingerprint, etc. Likewise, those recognition performance metrics are not suitable for the evaluation of protection performance. In \cite{Kuan:2005ut}, irreversibility is discussed theoretically and experimentally. Their experimental analysis is similar to  \cite{Jin:2004kc}. In their theoretical analysis, irreversibility is defined as the complexity of finding an exact original biometric feature vector from a transformed template and its corresponding key. BioHash is a lossy function, hence it satisfies their notion of irreversibility with some security parameter. However, in the real situation, the adversary usually does not have to find an exact original biometric feature, but enough to find an biometric feature which can be accepted by the biometric authentication system. The latter is trivially easy, given a transformed template and its corresponding key, randomly chosen biometric features will be accepted with probability FAR. Thus, more realistic notion of irreversibility is required.

\subsection{Biometric cryptosystem}
Biometric cryptosysm refers to a series of research motivated by fuzzy commitment and fuzzy vault proposed by Juels and Watenburg\cite{Juels:1999kz} and Juels and Sudan\cite{Juels:2002hd} respectively. Instead of applying sophisticated feature extraction and matching algorithms, they abstracted the metric space of biometrics matching as a hamming distance or a set difference respectively, and make use of error-correcting codes to check if the distance of two biometric features are within a correctable range. Dodis, Reyzin and Smith\cite{Dodis:2004uh} generalized them to secure sketch covering any {\it transitive} metric space, that is, a metric space $\cal M$ has a family of permutations $\pi \in \Pi$ such that $\Pi$ is distance preserving: $d(a,b)=d(\pi(a),\pi(b))$ and for any two elements $a,b \in \cal M$ there exists $\pi_i \in \Pi$: $\pi_i(a)=b$. 

None of them conducted experimental analysis both on  {\it recognition performance} (Section \ref{sect:RecPerf}) and {\it protection performance} (Section \ref{sect:ProtectPerf}). Rather, {\it irreversibility} (Section \ref{sect:Irr}) for their {\it un-keyed} schemes are theoretically analyzed. They demonstrated that fuzzy schemes have strong {\it irreversibility} in a practical parameter setting, but introduced impractical assumptions. As shown in this paper, any {\it un-keyed} schemes cannot satisfy {\it irreversibility} in a practical setting for a biometrics application (see Theorem \ref{thm:UNARCH-IRR} in this paper). Those impractical assumptions are considered essential in the analysis. Namely, Juels and Watenburg\cite{Juels:1999kz} assumes uniform distribution on biometric features, and Juels and Sudan\cite{Juels:2002hd} does not assume uniform distribution on elements in a set whereas assumes elements in a set are chosen independently. Dodis, Reyzin and Smith\cite{Dodis:2004uh} evaluated  {\it irreversibility} of secure sketch and fuzzy extractor with a general distribution on biometric feature, hence falls to insecure with a practical parameter setting for biometrics applications.

Sutcu, Li and Memon\cite{Sutcu:2007um} applied a secure sketch\cite{Dodis:2004uh} to a face recognition system, and measured {\it biometric performance} and estimated a lower-bound of {\it irreversibility}. They reported degradation of recognition performance introduced by secure sketch was negligible, but the lower-bound of complexity to break {\it irreversibility} was barely $20$ bits. Arakala, Jeffers and Horadam\cite{Arakala:2007vv} and Chang and Roy\cite{Chang:2007wz} applied to fingerprint recognition system and reported similar results.

\subsection{Related Security Metrics}
As we have seen until now, there are two separate line of research, and there exits a gap in the way of evaluation of  {\it recognition performance} and {\it protection performance} between feature transformation approach and biometric cryptosystem. 
Thus, relations of security statements were ambiguous, and it was not easy to compare the security of proposed schemes. Recently, there are attempts to try to unify the evaluation methods and give metrics applicable to all biometric template protection schemes. 

Nagar, Nandakumar and Jain\cite{Nagar:2010tg} proposed such security metrics. Their security metrics consists of six items: $\mathtt{FAR}_\mathtt{UK}$, $\mathtt{FAR}_\mathtt{KK}$, $\mathtt{IRIS}$, $\mathtt{IRID}$, $\mathtt{CMR}_\mathtt{T}$ and $\mathtt{CMR}_\mathtt{O}$. The first two items exactly correspond to our proposal, {\it accuracy} and {\it biometric performance}.  $\mathtt{IRIS}$, the Intrusion Rate due to Inversion for the Same biometric system, and $\mathtt{IRID}$, the Intrusion Rate due to Inversion for a Different biometric system, are related to our metric of {\it $\epsilon$-$\{PI, AD\}$-pseudo-authorized leakage irreversibility} in Definition \ref{def:PALIRR}. Our metric gives the upper-bound of the intrusion probability for all probabilistic polynomial-time inverters, whereas $\mathtt{IRIS}$ and $\mathtt{IRID}$ give the intrusion probability for the best possible inverter. $\mathtt{IRIS}$ and $\mathtt{IRID}$ can be evaluated experimentally, hence suitable metrics for algorithms in the feature transformation approach. However, $\mathtt{IRIS}$ and $\mathtt{IRID}$ should be considered that it gives the lower assurance in {\it irreversibility}, as far as there is no evidence that the best possible inverter used in the evaluation is the best of all probabilistic polynomial-time inverters. Similarly,  $\mathtt{CMR}_\mathtt{T}$, the Cross Match Rates in the Transformed feature domain, and $\mathtt{CMR}_\mathtt{O}$, the Cross Match Rates in the Original feature domain, are related to our {\it diversity} and {\it $\epsilon$-$\{PI, AD\}$-unlinkability}, respectively in Definition \ref{def:UNLINK}. 

Simoens, Yang, Zhou, Beato, Busch, Newton and Preneel\cite{SYZBBNP2012} proposed nearly all aspect of requirements normally expected to template protection algorithms, namely from technical performance such as recognition accuracy, throughput and storage requirement, protection performance through operational performance. Based on their proposal,  this paper focuses on the formal definitions of the recognition performance and the protection performance for precise discussions. For recognition performance,  their {\it accuracy}\cite{SYZBBNP2012} and {\it diversity}\cite{SYZBBNP2012} exactly corresponds to our {\it biometric performance} and {\it diversity}. Further, we introduced another {\it accuracy} which corresponds to $\mathtt{FAR}_\mathtt{UK}$ in Nagar et al.\cite{Nagar:2010tg} to demonstrate the performance advantage of two-factor template protection algorithms. For protection performance, their {\it irreversibility}\cite{SYZBBNP2012}  and {\it unlinkability}\cite{SYZBBNP2012} exactly corresponds to ours. {\it Irreversibility}\cite{SYZBBNP2012}  is further divided into {\it full-leakage irreversibility}, {\it authorized-leakage irreversibility} and {\it pseudo-aurhorized-leakage irreversibility} depending on the differences of goals for adversary. These three notions of {\it irreversibility} is formally defined and discussed their relations in Section \ref{sect:Irr}. {\it Unlinkability}\cite{SYZBBNP2012} is defined as the false cross match rate ($\FCMR$) and the false non-cross match rate ($\FNCMR$). These rate is measured as the performance of a {\it cross-comparator}. Similarly, if one could give an upper-bound of $\FCMR$ and $\FNCMR$ for all probabilistic polynomial-time  {\it cross-comparator}, then {\it unlinkability} can be theoretically evaluated with the high assurance level. On the other hand, if these rates are given experimentally for the best possible  {\it cross-comparator},  {\it unlinkability} is evaluated with lower assurance level. These are discussed in more detail in Section \ref{sect:Unlink}.

\section{Preliminaries}\label{sect:prelim}
In this section, we will explicitly formulate 
{\em biometric template protection $($BTP$)$ algorithms}. 
In this paper, we discuss BTP algorithms utilizing 
a common modality and a common feature extraction algorithm. 
Namely we do not discuss BTP algorithms using multi-biometrics. 
\par
Let $\U$ be a finite set consisting 
of all users who have biometric characteristics utilized in BTP algorithms. 
Assume that each user $u\in\U$ has 
his/her own biometric characteristic $b_u$ and therefore, 
in the following, we identify $u$ with $b_u$ and 
use the notation $u$ instead of $b_u$, 
namely, the set $\U$ can be regarded as a set consisting of 
all individuals' biometric characteristics. 
A biometric recognition system captures 
biometric samples from biometric characteristics presented 
to the sensor of the system, extracts 
biometric features from biometric samples, and 
verifies or identifies users by using their biometric features. 
We assume that each user's biometric features 
are represented as a digital element 
$x\in\M$ of a finite set $\M$. 
We call $x$ a {\em feature element} of $u$. 
Since two feature elements generated from $u$ are rarely identical, 
we let $X_u$ denote a random variable 
on $\M$ representing noisy variations of 
feature elements of $u$, namely 
$P(X_u=x)$ is the probability that 
a biometric sample of captured from $u$ will be 
represented as $x$. 
Let ${\mathbf R}$ be the set of all real numbers and 
let $d:\M\times\M\to{\mathbf R}$ be a {\em 
semimetric function} on $\M$, namely the real-valued function $d$ satisfies 
the following three conditions:
\begin{align*}
(\i)\quad& d(x,y)\geq0\\
(\ii)\quad& d(x,y)=0\text{ if and only if }\;x=y\\
(\iii)\quad& d(x,y)=d(y,x)
\end{align*}
for all $x,y\in\M$. 
Then $\M$ is called a {\em semimetric space} associated with $d$. 
For any $x\in\M$, $\M_{\tau}(x)=\{x'\;|\;d(x,x')\leq\tau\}$ is called 
the {\em $\tau$-neighborhood of $x$}. 
Let $f$ be an algorithm (or a function) on $\M$ whose 
input $x\in\M$ is chosen according to a random variable $X$. 
Let $f(X)$ denote a random variable induced on the image of $f$. 
For any set $T$, the notation $t\randomchosen T$ 
denotes that $t$ is chosen from the set $T$ uniformly at random. 
For any random variable $X$ on a set $\M$, the notation 
$x\leftarrow X$ 
denotes that $x$ is chosen according to $X$. 
For any function $f$ on the set $\M$, 
the notation 
$\mathop{\E}\limits_{x\leftarrow X}f(x)$ denotes the expected value 
of $f$ under the condition that $x$ 
is chosen according to the random variable 
$X$, namely 
\[
\mathop{\E}\limits_{x\leftarrow X}f(x)
=\sum\limits_{x\in\M}Pr[X=x]f(x)\;.
\]
In particular, 
\begin{align*}
\mathop{\E}_{x\leftarrow X}\Pr[\text{an event of $x$}]
=&\sum\limits_{x\in\M}\Pr[X=x]\Pr[\text{an event of $x$}\;|\;X=x]\\
=&\sum\limits_{x\in\M}\Pr[X=x, \;\text{an event of $x$}]\;.
\end{align*}
Traditional biometric comparison algorithms are 
assumed to utilize an ordinary comparison method 
which, for an enrolled feature element and 
a freshly extracted feature element $x'$ during verification, 
decides $\match$ if $d(x,x')\leq\tau$, 
and otherwise $\nonmatch$ by using a decision threshold $\tau$. 
Then, the {\em false non-match rate} $\FNMR_{d\leq\tau}$ and 
the {\em false match rate} $\FMR_{d\leq\tau}$ 
are formulated as follows: 
\begin{align}\label{eqn:FNMRandFMR}
&\FNMR_{d\leq\tau}=
\mathop{\E}_{
{\scriptsize
\begin{array}{l}
u\randomchosen \U\\
x,x'\randomchosen X_u
\end{array}}}
\Pr\left[d(x,x')>\tau\right]\\
&\FMR_{d\leq\tau}=
\mathop{\E}_{
{\scriptsize
\begin{array}{l}
(u,v)\randomchosen(\U\times\U)^{\diff}\\
x\randomchosen X_u, y\randomchosen X_v
\end{array}}}
\Pr\left[d(x,y)\leq\tau\right]\;.\nonumber
\end{align}
where $(\U\times\U)^{\diff}=\{(u,v)\in\U\times\U\;|\;u\neq v\}$ and 
$\#\U$ denotes the number of elements of $\U$. 
\paragraph{Biometric template protection algorithms}
We will give a explicit formulation of 
biometric template protection (BTP) algorithms as follows. 
\begin{definition}[BTP algorithms]\label{bidscheme}
A biometric template protection $($BTP$)$ algorithm $\Pi$ 
is a tuple of polynomial-time algorithms $\Gen$, $\PIE$, 
$\PIR$, $\PIC$, namely 
$\Pi=(\Gen,\PIE,\PIR,\PIC)$. 
Let $\Gen$ is an algorithm which on input $1^k$ returns 
a finite set $\U$ of biometric characteristics, the associated 
random variables $X_u$, $u\in\U$, over a semimetric space $\M$, and 
the public parameters $\params$, where $k$ is a security parameter. 
Let $\PIE$ be a randomized algorithm which on input $x\in\M$ 
returns a pair $(\pi,\alpha)$ of two data 
$\pi\in\M_{\PI}$ and $\alpha\in\M_{\AD}$, where 
$\M_{\AD}$ are finite sets. 
The algorithm 
$\PIE$ is called a {\em pseudonymous identifier encoder}. 
The first output $\pi$ $($resp. the second output $\alpha$ 
of $\PIE$ is called a {\em pseudonymous identifier $($PI$)$ 
for enrollment} $($resp. {\em auxiliary data $($AD$)$}$)$ and is denoted by 
$\pi=\PIE_1(x)$ $($resp. $\alpha=\PIE_2(x)$$)$. 
The algorithm $\PIE$ can be regarded as a pair of 
two randomized algorithms $\PIE_1$ and $\PIE_2$. 
\par
In the enrollment phase, 
a biometric characteristic $u\in\U$ is 
submitted to the system, 
a feature element $x\in\M$ 
is generated according to the distribution $X_u$, 
$\PIE$ outputs $(\pi,\alpha)$ on input $x$, 
and $\pi$ and $\alpha$ are stored in storages. 
Note that $\pi$ and $\alpha$ are not necessarily 
stored together in the same storage. 
\par
Let $\PIR$ be a deterministic algorithm which, 
on input $\alpha\in\M_{\AD}$ and $x'\in\M$, 
returns a data $\pi'\in\M'_{\PI}$ for verification, 
where $\M'_{\PI}$ is a finite set. 
The data $\pi'=\PIR(\alpha,x')$ is called 
a {\em pseudonymous identifier for verification}. 
Let $\PIC$ be a deterministic algorithm which, 
on input $\pi\in\M_{\PI}$ and $\pi'\in\M'_{\PI}$, returns 
either $\match$ or $\nonmatch$. 
The algorithms 
$\PIR$ and $\PIC$ are called a {\em pseudonymous identifier recorder} and 
a {\em pseudonymous identifier comparator}, respectively. 
\par
In the verification phase, 
a biometric characteristic $u\in\U$ is freshly presented to the system, 
a new feature element $x'\in\M$ is generated according to $X_u$. 
The verification entity receives a PI $\pi$, 
an AD $\alpha$ and $x'$, computes $\pi'=\PIR(\alpha,x')$, and 
outputs $\PIC(\pi,\pi')\in\{\match,\; \nonmatch\}$. 
\end{definition}
Note that the terms, pseudonymous identifier (PI), 
auxiliary data (AD), are defined 
in ISO/IEC 24745 \cite{ISO24745} (cf. \cite{SYZBBNP2012}). 
A pseudonymous identifier (PI) is defined to be 
a set of data that represents an individual or data subject within a certain domain by means of a protected identity 
and is used as a reference for verification 
by means of a captured biometric sample and auxiliary data. 
It is desirable that the PI does not allow the retrieval of the enrolled 
biometric feature element and multiple ``unlinkable'' PI's can be derived from the same biometric characteristic. 
Auxiliary data (AD) is defined to be a set of data 
that can be required to reconstruct 
pseudonymous identifiers during verification. 
In some scheme, AD depends on the enrolled biometric feature element. 
\par
A pair $(\pi,\alpha)$ 
of PI and AD is called a {\em protected template $($PT$)$} in \cite{SYZBBNP2012} or a {\em renewable biometric reference} in \cite{ISO24745}. 
In \cite{SYZBBNP2012}, in general, PTs are assumed to be public. 
However, most existing BTP algorithms require secrecy of PT. 
Because, in the real world, for some modalities (e.g. fingerprint, iris, face and so on) there are many public large databases, and therefore, 
the adversary can find a matching sample by entirely running such a database against a stolen PT. 
Therefore, in this paper, 
both PIs and ADs are assumed to be secret information. 
Each user's PI and AD are separately stored in different storages, 
for example, in application to $2$-factor authentication systems, 
every PI is stored together with each user's ID 
in the database and each user's AD is stored in the user's smart card. 
We will discuss the recognition performance and the security performance when one of (or both) PI and AD is leaked. 
Simoens et al. \cite{SYZBBNP2012} regard such a data separation 
as an additional property of BTP. 
\begin{definition}[2-factor BTP]
We will define a $2$-factor BTP authentication algorithms 
in which a biometric characteristic is the first authentication factor. 
There are two possibilities from the viewpoint of data separation. 
A scheme which utilizes ADs as second factors and 
stores PIs for verification in the database is called 
a {\em $\AD$-2-factor BTP}. 
Reversely, a scheme which utilizes PIs for verification 
as second factors and stores ADs in the database is called 
a {\em $\PI$-2-factor BTP}. 
\end{definition}

\section{Recognition performance for BTP algorithms}\label{sect:RecPerf}
In this section, 
we especially focus on recognition performance 
as technical performance of BTP algorithms $\Pi$. 
For the simplicity, we will fix a security parameter $k$. 
Therefore, a set $\U$ of biometric characteristics, 
the associated random variables $X_u$, $u\in\U$, and 
the public parameters $\params$ are fixed. 
\subsection{Accuracy}\label{sect:Accuracy}
For any biometric template protection $($BTP$)$ algorithm 
$\Pi=(\PIE,\PIR,\PIC)$, 
the false non-match rate of $\Pi$, $\FNMR_{\Pi}$, 
is the probability that 
a mated pair of PT and biometric sample are falsely 
declared to be $\nonmatch$, namely, 
\begin{align*}
\FNMR_{\Pi}
=&
\mathop{\E}_
{\scriptsize
\begin{array}{l}
u\randomchosen\U\\
x\leftarrow X_u\\
(\pi,\alpha)\leftarrow\PIE(X_u)
\end{array}}
\Pr\left[
\begin{array}{c}
\PIC\bigl(\pi,\PIR(\alpha,x)\bigr)=\nonmatch
\end{array}\right]
\end{align*}
Here, we will define 
recognition accuracy metrics for $2$-factor BTPs, 
which are called {\em total performance} and 
naturally introduced from the notion, 
data separation, discussed 
by Simoens et al. \cite[Section 4.4]{SYZBBNP2012}. 
The {\em false match rate for total performance} 
of $\AD$-$2$-factor BTP $($resp. $\PI$-2-factor BTP$)$ $\Pi$, 
$\FMR_{\Pi,\;\AD}^{\ToPerf}$ 
(resp. $\FMR_{\Pi,\;\PI}^{\ToPerf}$), 
is the probability that a zero-effort impostor's presentation 
of his own biometric characteristic $u\in\U$ 
along with a 2nd factor $\alpha\in\M_{\AD}$ (resp. $\pi\in\M_{\PI}$) 
generated from $u$ 
is falsely declared to match a non-mated reference data 
$\pi\in\M_{\PI}$ (resp. $\alpha\in\M_{\AD}$) generated from 
a biometric characteristic $v\in\U\bs\{u\}$. 
The metrics $\FMR_{\Pi,\;\AD}^{\ToPerf}$ and 
$\FMR_{\Pi,\;\PI}^{\ToPerf}$ are respectively formulated by 
\begin{align*}
\FMR_{\Pi,\;\AD}^{\ToPerf}
=&
\mathop{\E}_{
{\scriptsize
\begin{array}{l}
(u,v)\randomchosen(\U\times\U)^{\diff}\\
x\leftarrow X_u\\
(\pi,\alpha)\leftarrow\PIE(X_u)\\
(\pi',\alpha')\leftarrow\PIE(X_v)
\end{array}}
}
\Pr\left[
\begin{array}{c}
\PIC\bigl(\pi',\PIR(\alpha,x)\bigr)=\match
\end{array}\right]\\
\FMR_{\Pi,\;\PI}^{\ToPerf}
=&
\mathop{\E}_{
{\scriptsize
\begin{array}{l}
(u,v)\randomchosen(\U\times\U)^{\diff}\\
x\leftarrow X_u\\
(\pi,\alpha)\leftarrow\PIE(X_u)\\
(\pi',\alpha')\leftarrow\PIE(X_v)
\end{array}}
}
\Pr\left[
\begin{array}{c}
\PIC\bigl(\pi,\PIR(\alpha',x)\bigr)=\match
\end{array}\right]\;.
\end{align*}
Nagar et al. \cite{Nagar2010} propose these metrics 
as the false accept rate with unknown transformation parameters, 
$\FAR_{UK}$. 
\par
By measuring the above metrics, 
$\FNMR_{\Pi}$, 
$\FMR_{\Pi,\;\AD}^{\ToPerf}$, and 
$\FMR_{\Pi,\;\PI}^{\ToPerf}$, 
we can totally evaluate the recognition performance of $2$-factor BTPs. 
However, a $2$-factor BTP can achieve a high recognition performance 
when the recognition accuracy contributed by one factor is high, 
even if the recognition accuracy contributed by the other factor is poor. 
Therefore, we need to evaluate the recognition accuracy 
achieved only by using one factor. 
In the following sections, 
Section \ref{sect:BioPerf} and \ref{sect:Div}, 
we will define metrics for such recognition accuracy. 

\subsection{Biometric Performance}\label{sect:BioPerf}
In this section, we will define a metric 
for the recognition accuracy achieved only by the 1st factor, biometrics. 
The {\em false match rate for biometric performance} of $\Pi$, 
$\FMR_{\Pi}^{\BioPerf}$, is the probability that 
a zero-effort impostor's presentation 
of his own biometric characteristic $u\in\U$ 
along with a correct $2$nd factor is falsely 
declared to match a genuine reference data. 
Then the metric $\FMR_{\Pi}^{\BioPerf}$ is formulated by 
\begin{align*}
\FMR_{\Pi}^{\BioPerf}
=&
\mathop{\E}_{
{\scriptsize
\begin{array}{l}
(u,v)\randomchosen(\U\times\U)^{\diff}\\
x\leftarrow X_u\\
(\pi,\alpha)\leftarrow\PIE(X_v)\\
\end{array}}}
\Pr\left[\PIC(\pi,\PIR(\alpha,x))=\match\right]\;.
\end{align*}
Simoens et al. \cite{SYZBBNP2012} discussed this metric as an ordinary 
recognition accuracy metric, the false match rate, because they mainly 
consider biometric-based single factor authentication systems which 
stores PIs and ADs in the database. 
Moreover, Nagar et al. \cite{Nagar2010} propose this metric as 
{\em the false accept rate with known transformation parameters}, 
$\FAR_{KK}$. 
\par
This metric can be regarded as a 
metric for security against 
impersonation when a user's 2nd factor is leaked. 
In the above notion, biometric performance, 
the adversary assumed to be very weak, namely 
he presents his own biometric characteristic along 
with obtained genuine user's $2$nd factor. 
However, in order to strictly evaluate security against impersonation, 
we need to define a stronger attack model. 
We would discuss such a rigorous security 
in another paper in preparation. 

\subsection{Diversity}\label{sect:Div}
Diversity is the notion 
which ensures renewability for $2$-factor BTPs. 
Namely, after a PT generated from $u\in\U$ is renewed, 
a presentation of $u$ along with the old 2nd factor should 
not be declared to match the new reference data. 
Diversity is also the property that PTs should not allow cross-matching 
across databases in different authentication systems. 
(cf. \cite[\III]{Jain:2008vv}, \cite[Sect. 3.3]{Nagar2010}, 
\cite[Sect. 3.5]{SYZBBNP2012}). 
We will define a metric for diversity as follows. 
The {\em false match rate for diversity} 
of BTP algorithm $\Pi$, $\FMR_{\Pi}^{\Div}$, 
is the probability that a presentation of 
a biometric characteristic $u\in\U$ along 
with a 2nd factor generated from $u$ 
is falsely declared to match a new reference data 
freshly generated from the same $u$. 
The metrics $\FMR_{\Pi}^{\Div}$ is formulated by 
\begin{align*}
\FMR_{\Pi}^{\Div}
=\mathop{\E}_{
{\scriptsize
\begin{array}{l}
u\randomchosen\U\\
x\leftarrow X_u\\
(\pi,\alpha)\leftarrow\PIE(X_u)\\
(\pi',\alpha')\leftarrow\PIE(X_u)
\end{array}}
}
\Pr\left[
\begin{array}{c}
\PIC\bigl(\pi,\PIR(\alpha',x)\bigr)=\match
\end{array}\right]\;.
\end{align*}
Nagar et al. \cite{Nagar2010} proposes 
this metric as the {\em cross match rate}, $\CMR$. 
Here we consider the corresponding entropy $H=-\log\FMR_{\Pi}^{\Div}$. 
Then, it indicates that the distribution of 
PTs generated form a biometric characteristic are almost 
the same as the uniform distribution on $H$-bit binary strings, namely 
$2^H$ independent PTs can be generated 
from a biometric characteristic. 
Simoens et al. \cite{SYZBBNP2012} propose 
the number of such ``independent'' PTs as a metric for diversity. 
\par
Diversity can be regarded as a metric for security 
against impersonation 
when a user's biometric characteristic is leaked. 
For example, in the real world, 
a fingerprint left on a glass is abused by a malicious user. 
However, in the above diversity notion, the adversary assumed to be 
very weak, namely he submits a $2$nd factor 
randomly generated from the obtained biometric characteristic. 
By using the obtained biometric characteristic, 
a stronger adversary might be able to 
find a $2$nd factor which makes $\PIC$ return $\match$ 
with extremely higher probability. 
We would discuss such a strict security notion 
in another paper in preparation. 

\section{Protection peformance for BTP algorithms}\label{sect:ProtectPerf}
\subsection{Irreversibility}\label{sect:Irr}
Suppose that 
the adversary obtains (a part of) a PT leaked from 
the database or from the user's storage devices. 
The adversary might be able to 
recover a feature element close to the original feature element from which 
the PT is generated. 
Form the recovered feature element, 
he might create a physical spoof of the user's biometric characteristic 
and impersonate the user by presenting the fake biometric characteristic 
to the system. 
Irreversibility is a requirement that 
it should be hard to recover an original feature element 
(or a neighborhood of it) from (a part of) a PT, 
which ensures the security in the case of leakage of PTs. 
\par
For each nonempty subset $\Lambda\neq\phi$ of the terms $\{PI,AD\}$ 
and any PT $(\pi,\alpha)$, 
let $(\pi,\alpha)_{\Lambda}$ denote a subset 
of $\{\pi,\alpha\}$ defined by $(\pi,\alpha)_{\{PI,AD\}}=(\pi,\alpha)$, 
$(\pi,\alpha)_{\{PI\}}=\pi$, and 
$(\pi,\alpha)_{\{AD\}}=\alpha$. 
We call $(\pi,\alpha)_{\Lambda}$ a {\em $\Lambda$-subset} of $(\pi,\alpha)$
\par
We will define a {\em irreversibility game $($IRR Game$)$} 
between the challenger $\Ch$ and the adversary $\Adv=(\Adv_1,\Adv_2)$, where 
$\Adv_1$ is a probabilistic polynomial-time (ppt) adversary which 
is given the algorithms and the parameters of $\Pi$ and sends a state 
to $\Adv_2$, and $\Adv_2$ is a ppt adversary who is given a $\Lambda$-subset 
of a PT generated from an feature element $x\in\M$ extracted 
from a randomly chosen biometric characteristic and 
attempts to guess (a neighborhood of) the original feature element $x$. 
\par
Recently, for most major modalities, 
there are many databases available to the public. 
Therefore, it is natural to assume that 
the adversary easily obtains a huge database of biometric samples. 
In this case, the adversary can 
performs an offline attack and successfully find a target feature element. 
In order to formulate such a practical situation, we will define 
an oracle from which the adversary can obtain feature elements corresponding 
to biometric characteristics submitted as queries. 
More precisely, let $\Samp$ be an oracle which, on input $u\in\U$, 
chooses $x\in\M$ according to $X_u$ and returns $x$. 
We assume that the challenger and the adversary are 
allowed to make polynomial-time queries to $\Samp$ 
before he returns his guess. 
\par
For any subset $\phi\neq\Lambda\subset\{\PI,\AD\}$ and 
any real number $\tau\geq0$, 
we define {\em $\Lam$-$\tau$-authorized leakage game 
$($$\Lam$-AL$_{\tau}$ IRR Game$)$} 
(resp. {\em $\Lam$-pseudo authorized leakage game 
$($$\Lam$-PAL IRR Game$)$}) as follows. 
\vskip.2cm
\noindent
{\bf $\Lam$-AL$_{\tau}$ IRR Game} (resp. {\bf $\Lam$-PAL IRR Game})
\begin{description}
\item[Step 1.] The challenger $\Ch$ inputs $1^k$ into $\Gen$ and 
$\Gen$ returns $\U$, $X_u$, $u\in\U$, and the parameters $\params$. 
The challenger $\Ch$ sends $(\params,\Lam,\tau)$ (resp. 
$(\params,\Lam)$) to the adversary $\Adv_1$. 
\item[Step 2.] The adversary $\Adv_1$ receives $(\params,\Lam,\tau)$ 
(resp. $(\params,\Lam)$) and sends a state $s$ to $\Adv_2$. 
The adversary $\Adv_1$ is allowed to make polynomial-time queries to $\Samp$ 
before he sends $s$ to $\Adv_2$. 
\item[Step 3.] The challenger $\Ch$ chooses a biometric characteristic 
$u\in\U$ uniformly at random, submits $u$ to the sampling oracle $\Samp$, and 
gets a feature element $x\in\M$ as an answer from $\Samp$. 
The challenger $\Ch$ inputs the feature element $x$ into $\PIE$, 
gets the output $(\pi,\alpha)$, and sends $(\pi,\alpha)_\Lam$ 
to the adversary 
$\Adv_2$. 
\item[Step 4.] The adversary $\Adv_2$ receives the state $s$ and 
$(\pi,\alpha)_\Lam$ from $\Adv_1$ and $\Ch$, respectively, 
and returns $x'\in\M$. The adversary $\Adv_2$ is allowed to make polynomial-time queries to $\Samp$ before he returns his guess. 
\end{description}
If $d(x,x')\leq\tau$ (resp. $\PIC(\pi,\PIR(\alpha,x'))=\match$), 
then the adversary $\Adv=(\Adv_1,\Adv_2)$ wins. 
\par
Traditional biometric recognition algorithms, 
which do not use BTP algorithms, determines decision thresholds $\tau$ 
to minimize the false non-match rate $\FNMR_{d\leq\tau}$ or 
the false match rate $\FMR_{d\leq\tau}$ (cf. (\ref{eqn:FNMRandFMR})). 
If the adversary obtains a $\Lam$-subset of a PT and 
successfully recovers a feature element close to the original feature element, 
then he can impersonate the user in traditional authentication systems. 
Since some BTP algorithms might accept feature elements outside 
the $\tau$-neighborhood of the original feature element, 
the adversary in $\Lam$-PAL IRR Game might 
find a feature element $x'$ such that 
$\PIC(\pi,\PIR(\alpha,x'))=\match$ but $d(x,x')>\tau$. 
\par
For any feature element $x\in\M$, 
{\em the match rate of the feature element $x$ 
with respect to $d\leq\tau$} (resp. 
{\em the reverse match rate of the feature element $x$}) 
$\MR_{d\leq\tau}(x)$ (resp. $\rMR_{\Pi}(x)$ ) 
is the probability 
that a feature element $x'\in\M$ (resp. a PT $(\pi,\alpha)$) 
generated from a randomly chosen biometric characteristic $u\in\U$ 
satisfies $d(x,x')\leq\tau$ (resp. $\PIC(\pi,\PIR(\alpha,x))=\match$), 
which is formulated by 
\begin{align}
&\MR_{d\leq\tau}(x)=
\mathop{\E}_{
{\scriptsize
\begin{array}{l}
x'\leftarrow X(\U)
\end{array}}
}
\Pr\left[
\begin{array}{c}
d(x,x')\leq\tau
\end{array}\right]\label{eqn:MR_AL}\\
&\rMR_{\Pi}(x)=
\mathop{\E}_{
{\scriptsize
\begin{array}{l}
(\pi,\alpha)\leftarrow\PIE(X(\U))
\end{array}}
}
\Pr\left[
\begin{array}{c}
\PIC(\pi,\PIR(\alpha,x))=\match
\end{array}\right]\;.\label{eqn:rMR_PAL}
\end{align}
Put $m_{d\leq\tau}=\mathop{\max}\limits_{x}\MR_{d\leq\tau}(x)$ and 
$m_{\Pi}=\mathop{\max}\limits_{x}\rMR_{\Pi}(x)$. 
In $\Lam$-AL$_{\tau}$ IRR Game (resp. $\Lam$-PAL IRR Game), 
the optimal strategy of an adversary $\Adv'$
who is not given $(\pi,\alpha)_\Lam$ is to return 
a feature element $x$ satisfying 
$\MR_{d\leq\tau}(x)=m_{d\leq\tau}$ (resp. $\rMR_{\Pi}(x)=m_{\Pi}$) 
and then the success probability of the adversary $\Adv'$ is 
equals to $m_{d\leq\tau}$ (resp. $m_{\Pi}$). 
Therefore, the advantage $\ADVANTAGE^{\text{\rm$\Lam$-AL$_{\tau}$ IRR}}_{\Pi,\Adv}$ 
(resp. $\ADVANTAGE^{\text{\rm$\Lam$-PAL IRR}}_{\Pi,\Adv}$) 
of the adversary $\Adv$ is defined by 
\begin{align*}
&\ADVANTAGE^{\text{\rm $\Lam$-AL$_{\tau}$ IRR}}_{\Pi,\Adv}
=\Pr[\text{$\Adv$ in $\Lam$-AL$_{\tau}$ IRR Game wins}]-m_{d\leq\tau}\\
&\ADVANTAGE^{\text{\rm $\Lam$-PAL IRR}}_{\Pi,\Adv}
=\Pr[\text{$\Adv$ in $\Lam$-PAL IRR Game wins}]-m_{\Pi}
\end{align*}
\begin{definition}[Authorized-leakage irreversibility (cf. \cite{SYZBBNP2012})] We say that a BTP algorithm $\Pi$ 
is {\em $\vep$-$\Lam$-$\tau$-authorized-leakage irreversible 
$($$\vep$-$\Lam$-AL$_{\tau}$ IRR$)$} if 
$\ADVANTAGE^{\text{\rm $\Lam$-AL$_{\tau}$ IRR}}_{\Pi,\Adv}<\vep$ 
for any ppt adversary $\Adv$. 
In particular, we say that $\Pi$ 
is {\em $\vep$-$\Lam$-full-leakage irreversible 
$($$\vep$-$\Lam$-FL IRR$)$} if 
$\ADVANTAGE^{\text{\rm $\Lam$-AL$_{0}$ IRR}}_{\Pi,\Adv}<\vep$ 
for any ppt adversary $\Adv$. 
\end{definition}
\begin{definition}[Pseudo-authorized-leakage irreversibility (cf. \cite{SYZBBNP2012})]\label{def:PALIRR}
We say that a BTP algorithm $\Pi$ 
is {\em $\vep$-$\Lam$-pseudo-authorized-leakage irreversible 
($\vep$-$\Lam$-PAL IRR)} 
if $\ADVANTAGE^{\text{\rm $\Lam$-PAL IRR}}_{\Pi,\Adv}<\vep$ 
for any ppt adversary $\Adv$. 
\end{definition}
The above definitions immediately implies the following theorem. 
We omit the proof. 
\begin{theorem}\label{thm:IRR}
Fix any nonempty subset $\Lam\subset\{PI,AD\}$ and 
any real numbers $\vep>0$ and $\tau\geq0$. 
If a BTP algorithm $\Pi$ is $\vep$-$\Lam$-AL$_{\tau}$ IRR, 
then $\Pi$ is $(\vep+m_{d\leq\tau}-m_{d\leq0})$-$\Lam$-FL IRR. 
\par
Moreover, assume that 
$\tau$ satisfies the condition that, 
for any $x\in\M$ and any 
PT $(\pi,\alpha)$ generated from $x$, 
$\PIC(\pi,\PIR(\alpha,x'))=\match$ if $d(x,x')\leq \tau$. 
If a BTP algorithm $\Pi$ is $\vep$-$\Lam$-PAL IRR, 
then $\Pi$ is $(\vep+m_{Pi}-m_{d\leq\tau})$-$\Lam$-AL$_{\tau}$ IRR. 
\end{theorem}
Simoens et al. \cite{SYZBBNP2012} also introduce 
the above metrics, FL IRR, AL IRR, and PAL IRR, 
as the difficulty of determining 
(a neighborhood of) the original feature element. 
Note that, in the attack model in \cite{SYZBBNP2012} 
the adversary is given the whole PT. 
Here, we discuss unachievability of PAL IRR 
in the case when the adversary is given a whole PT, namely 
$\Lam=\{\PI,\AD\}$. 
Actually, when the adversary obtains both the PI and the AD, 
he can find a target feature element 
with extremely high probability 
by making a certain amount of queries to $\Samp$. 
We will more precisely discuss as follows. 
\par
For any PT $(\pi,\alpha)\in\M_{\PI}\times\M_{\AD}$, 
the {\em match rate of the PT $(\pi,\alpha)$}, 
$\MR_{\Pi}(\pi,\alpha)$, is the probability 
that a feature element $x'\in\M$ 
generated from a randomly chosen biometric characteristic $u\in\U$ 
satisfies $\PIC(\pi,\PIR(\alpha,x'))=\match$, 
which is formulated by 
\begin{align*}
\MR_{\Pi}(\pi,\alpha)
=&
\mathop{\E}_{
{\scriptsize
\begin{array}{c}
x'\leftarrow X(\U)
\end{array}}}
\Pr\left[\PIC(\pi,\PIR(\alpha,x'))=\match\right]\;.
\end{align*}
The $\MR_{\Pi}(\pi,\alpha)$ can be regarded 
as a random variable over the distribution of $(\pi,\alpha)$. 
Let $\MR_{\Pi}$ denote 
the average of $\MR_{\Pi}(\pi,\alpha)$, namely 
\begin{align*}
\MR_{\Pi}
=&
\mathop{\E}_{
{\scriptsize
\begin{array}{c}
(\pi,\alpha)\leftarrow\PIE(X(\U))
\end{array}}}
\MR_{\Pi}(\pi,\alpha)\;.
\end{align*}
Let $\sigma$ is the standard deviation of the 
$\MR_{\Pi}(\pi,\alpha)$. 
Then, from Chebyshev's inequality, we have 
\begin{align}\label{Cheby}
\Pr\left[\MR_{\Pi}(\pi,\alpha)>\MR_{\Pi}-
\frac{\sigma}{\sqrt{\delta}}\right]&\geq1-\delta
\end{align}
for any $\delta>0$. 
For the simplicity, we assume that 
$\MR_{\Pi}$ and $\sigma$ are constants 
independent of the security parameter $k$. 
Let $C$ be the variation coefficient of 
$\MR_{\Pi}(\pi,\alpha)$, namely 
$C=\disp\frac{\sigma}{\MR_{\Pi}}$. 
Assume that $C<1$. 
\begin{theorem}\label{thm:UNARCH-IRR}
For all $\vep<1-C^2-m_{\Pi}$, 
there exists no $\vep$-$\{\PI,\AD\}$-PAL IRR BTP algorithm.
\end{theorem}
In general, more accurate BTP algorithms $\Pi$ have smaller $C$. 
Therefore, Theorem \ref{thm:UNARCH-IRR} 
states that accurate BTP algorithms are unlikely to achieve 
sufficient irreversibility when both PI and AD are compromised. 
\par
We will prove Theorem \ref{thm:UNARCH-IRR} 
in Appendix \ref{sect:proofUNARCH-IRR}. 
We can also similarly prove unachievability of AL IRR when the adversary 
is given the whole PT under the 
assumptions slightly different from the case of PAL IRR. 
However, we omit a precise description of the statement and the proof 
in this e-print and will describe them in the full paper. 

\subsection{Unlinkability}\label{sect:Unlink}
For any nonempty subset $\Lam\subset\{\PI,\AD\}$, 
we will define $\Lam$-UNLINK Game between 
the challenger $\Ch$ and 
the adversary $\Adv=(\Adv_1,\Adv_2)$, where 
$\Adv$ is given $\Lam$-subsets of two PTs and attempts to 
guess whether 
the PTs are generated from the same biometric characteristics or not. 
In this game, $\Ch$ and $\Adv$ are allowed to make polynomial-time 
queries to the sampling oracle $\Samp$. 
\vskip.2cm
\noindent
{\bf $\Lam$-UNLINK Game}
\begin{description}
\item[Step 1.] The challenger $\Ch$ inputs $1^k$ into $\Gen$ and 
$\Gen$ returns $\U$, $X_u$, $u\in\U$, and the parameters $\params$. 
$\Ch$ sends $(\params,\Lam)$ to the adversary $\Adv_1$. 
\item[Step 2.] The adversary $\Adv_1$ receives 
$(\params,\Lam)$, outputs three feature elements 
$x$, $x_0$, $x_1$ depending on a distribution selected 
by $\Adv_1$, sends $(x,x_0,x_1)$ to $\Ch$, 
and sends a state $s$ to $\Adv_2$, where 
$s$ contains $(x,x_0,x_1)$. 
The adversary $\Adv_1$ is allowed to make polynomial-time queries to $\Samp$ 
before he sends $s$ to $\Adv_2$. 
\item[Step 3.] The challenger $\Ch$ flips 
the random coin $b\in\{0,1\}$, 
inputs $x$, $x_b$ into $\PIE$, gets 
$PT=\PIE(x)$ and 
$PT'=\PIE(x_b)$, and 
sends $(PT)_\Lam$ and $(PT')_\Lam$ to the adversary $\Adv_2$. 
\item[Step 4.] The adversary $\Adv_2$ receives 
the state $s$ and $(PT)_\Lam$ and $(PT')_\Lam$ from $\Adv_1$ and $\Ch$, 
and returns $b'\in\{0,1\}$ as a guess of $b$. 
The adversary $\Adv_2$ is allowed to make polynomial-time queries to $\Samp$ 
before he returns his guess. 
\end{description}
If $b'=b$, then the adversary $\Adv=(\Adv_1,\Adv_2)$ wins. 
The advantage $\ADVANTAGE^{\text{\rm $\Lam$-UNLINK}}_{\Pi,\Adv}$ 
of the adversary $\Adv$ over the random guess is formulated by 
\begin{align}\label{advantage_unlinkability}
\ADVANTAGE^{\text{\rm $\Lam$-UNLINK}}_{\Pi,\Adv}
&=\bigl|2\Pr\left[\Adv \;\text{wins}\right]
-1\bigr|
\end{align}
\begin{definition}[Unlinkability] \label{def:UNLINK}
We say that a BTP algorithm $\Pi$ is 
{\em $\vep$-$\Lam$-unlinkable $($$\vep$-$\Lam$-UNLINK$)$}, 
if $\ADVANTAGE^{\text{\rm$\Lam$-UNLINK}}_{\Pi,\Adv}<\vep$ 
for any ppt adversary $\Adv$. 
\end{definition}
Here, we will show unachievability of unlinkability when both PI and AD are compromised. 
\begin{theorem}\label{thm:UNARCH-UNLINK}
Assume that, for any $x\in\M$ and any PT $(\pi,\alpha)$ generated 
from $x$, 
$\PIC(\pi,\PIR(\alpha,x))=\match$. 
For any $\vep\leq 1-\MR_{\Pi}$, there exists no 
$\vep$-$\{\PI,\AD\}$-UNLINK BTP algorithm. 
\end{theorem}
In general, more accurate BTP algorithms $\Pi$ have smaller $\MR_{\Pi}$. 
Therefore, Theorem \ref{thm:UNARCH-UNLINK} 
states that accurate BTP algorithms are unlikely to achieve 
sufficient unlinkability when both PI and AD are compromised. 
\par
We will prove Theorem \ref{thm:UNARCH-UNLINK} 
in Appendix \ref{sect:proofUNARCH-IRR}. 
\par
Simoens et al. \cite{SYZBBNP2012} 
define a metric for unlinkability by using 
the {\em false cross match rate $($$\FCMR$$)$} and 
the {\em false non-cross-match rate $($$\FNCMR$$)$}. 
They define an adversary $\Adv^{cc}=(\Adv^{cc}_1,\Adv^{cc}_2)$, who is 
called the {\em cross-comparator}. 
In $\Lam$-UNLINK Game, 
the adversary $\Adv^{cc}_1$ chooses a pair $(u,v)\in(\U\times\U)^{\diff}$ 
of two different biometric characteristics, 
submits $u$ to $\Samp$ independently twice, and receives 
$x$ and $x_0$ respectively as the answers of two queries, 
moreover submits $v$ to $\Samp$, and receives $x_1$ as the answer, 
sends $(x,x_0,x_1)$ to $\Ch$, and sends a state 
$s$ containing $(x,x_0,x_1)$ to $\Adv^{cc}_2$. 
The adversary $\Adv^{cc}_2$ receives the state containing 
$(x,x_0,x_1)$ and $(PT_1)_\Lam$ and $(PT_2)_\Lam$ from 
$\Adv^{cc}_1$ and $\Ch$, respectively, 
and returns $b'\in\{0,1\}$ as a guess of $b$. 
\par
The false cross match rate $($$\FCMR$$)$ 
(resp. the false non-cross-match rate $($$\FNCMR$$)$) is the probability 
that, when $b=1$ (resp. $b=0$), 
the cross comparator $\Adv^{cc}$ 
falsely guesses that $b'=0$ (resp. $b'=1$), 
which is formulated as follows:
\begin{align*}
\FCMR^{\text{$\Lam$-UNLINK}}_{\Pi,\Adv^{cc}}&=
\mathop{\E}_{
{\footnotesize
\begin{array}{l}
(u,v)\randomchosen(\U\times\U)^{\diff}\\
(PT)_\Lam\leftarrow\PIE(X_u)\\
(PT')_\Lam\leftarrow\PIE(X_v)
\end{array}}}
\Pr[\text{$\Adv^{cc}$ returns $0$ in $\Lam$-UNLINK Game}]\\
\biggl(\text{resp.\;}
\FNCMR^{\text{$\Lam$-UNLINK}}_{\Pi,\Adv^{cc}}
&=
\mathop{\E}_{
{\footnotesize
\begin{array}{l}
u\randomchosen\U\\
(PT)_\Lam\leftarrow\PIE(X_u)\\
(PT')_\Lam\leftarrow\PIE(X_u)
\end{array}}}
\Pr[\text{$\Adv^{cc}$ returns $1$ in $\Lam$-UNLINK Game}]\biggr)\;.
\end{align*}
The advantage 
$\ADVANTAGE^{\text{$\Lam$-UNLINK}}_{\Pi,\Adv^{cc}}$ of 
the cross comparator 
can be interpreted as follows: 
\begin{align*}
\ADVANTAGE^{\text{$\Lam$-UNLINK}}_{\Pi,\Adv^{cc}}=
\bigl|1-\bigl(\,\FCMR^{\text{$\Lam$-UNLINK}}_{\Pi,\Adv^{cc}}
+\FNCMR^{\text{$\Lam$-UNLINK}}_{\Pi,\Adv^{cc}}\,\bigr)\bigr|\;.
\end{align*}

\section{Relations among security notions}
In this section, 
we will clarify relations among security notions, 
irreversibility and unlinkability, defined in the previous sections. 
We will prove that unlinkability 
is a stronger notion than authorized-leakage irreversibility. 
Therefore, unlinkability gives 
more rigorous assurance on privacy than irreversibility. 
Before describing the precise statement, 
we will prepare some notations. 
\par  
Let $P_{\tau}(x)$ be the probability that 
the $\tau$-neighborhood of $x'$ chosen according to the distribution 
$X(\U)$ has non-empty intersection 
with $\M_{\tau}(x)$, namely 
\begin{align*}
P_{\tau}(x)=\mathop{\E}_{
{\footnotesize
\begin{array}{l}
x'\leftarrow X(\U)
\end{array}}}
\Pr[\M_{\tau}(x)\cap\M_{\tau}(x')\neq\phi]\;.
\end{align*}
Note that, for any $\tau<\tau'$, $P_{\tau}(x)\leq P_{\tau'}(x)$. 
Put $p_{\tau}=\mathop{\max}\limits_{x}P_{\tau}(x)$ and 
$q_{\tau}=\mathop{\min}\limits_{x}P_{\tau}(x)$. 
Note that, for any $\tau<\tau'$, 
$p_{\tau}\leq p_{\tau'}$ and $q_{\tau}\leq q_{\tau'}$, and 
$q_0\leq \disp\frac{1}{\#\M}\leq p_0$ and 
the equality is attained if and only if $X(\U)$ is a uniform distribution. 
\begin{theorem}\label{thm:UNLINK-IRR} 
For any nonempty subset $\Lam\subset\{\PI,\AD\}$, 
if a BTP algorithm $\Pi$ is $\vep$-$\Lam$-UNLINK, then 
$\Pi$ is 
$\disp\frac{\vep+(p_{\tau}-q_{\tau})m_{d\leq\tau}}{1-p_{\tau}}$
-$\Lam$-AL$_{\tau}$ IRR 
for any $\tau\geq0$. 
\end{theorem}
We will prove Theorem \ref{thm:UNLINK-IRR} in 
Appendix \ref{sect:proofUNLINK-IRR}. 
\par
From Theorem \ref{thm:IRR} and Theorem \ref{thm:UNLINK-IRR}, 
we have the following figures, 
Figure \ref{RELS_PI_COMP} and Figure \ref{RELS_AD_COMP}, 
which indicate relations 
among irreversibility and unlinkability when 
$\Lam=\PI$ and $\Lam=\AD$, respectively. 
The notation $A\longrightarrow B$ means that 
the notion $A$ is stronger than the notion $B$. 
We avoide to show the figure in the case of $\Lam=\{\PI,\AD\}$, 
because, as mentioned after Theorem \ref{thm:UNARCH-IRR} and 
Theorem \ref{thm:UNARCH-UNLINK}, 
accurate BTP algorithms are unlikely to achieve 
sufficient irreversibility or unlinkability 
when both PI and AD are compromised. 
\begin{figure}[h]
\begin{minipage}{0.45\textwidth}
\begin{center}
$\begin{CD}
\text{PI-FL IRR} @.\\
@AAA @.\\
\text{PI-AL IRR}@<<<\text{PI-UNLINK}\\
@AAA @.\\
\text{PI-PAL IRR} @.
\end{CD}
$
\vskip.5cm
\caption{Relations among security notions 
when only PI is compromised.}\label{RELS_PI_COMP}
\end{center}
\end{minipage}
\hspace{.5cm}
\begin{minipage}{0.45\textwidth}
\begin{center}
$\begin{CD}
\text{AD-FL IRR}@.\\
@AAA @.\\
\text{AD-AL IRR}@<<<\text{AD-UNLINK}\\
@AAA @.\\
\text{AD-PAL IRR}@.
\end{CD}
$
\vskip.5cm
\caption{Relations among security notions 
when only AD is compromised.}\label{RELS_AD_COMP}
\end{center}
\end{minipage}
\end{figure}

\bibliographystyle{plain}	
\bibliography{RSM20120720}
\appendix
\section{Proof of Theorem \ref{thm:UNARCH-IRR}}\label{sect:proofUNARCH-IRR}
In this appendix, 
we will explicitly prove Theorem \ref{thm:UNARCH-IRR}. 
\vskip.2cm
\noindent
{\it Proof of Theorem $\ref{thm:UNARCH-IRR}$.} 
Put $\Lam=\{\PI,\AD\}$. 
We need to prove that, for any constant $\gamma$ satisfying 
$C^2<\gamma<1$, 
there exists an adversary $\Adv$ satisfying 
$\Pr[\text{$\Adv$ in $\Lam$-PAL IRR Game wins}]>1-\gamma$. 
We will define an adversary $\Adv_2$ 
who obtains a PT $(\pi,\alpha)$ from the challenger $\Ch$, 
makes polynomial-time queries 
to the sampling oracle $\Samp$, and returns 
a guess $x'\in\M$. 
\par
Fix a constant $\delta$ satisfying $C^2<\delta<\gamma$. 
Put $\mu=\MR_{\Pi}-\disp\frac{\sigma}{\sqrt{\delta}}$. 
Since $0<\mu<1$ and $\mu$ is a constant, 
there exists a constant number $N_{\delta}$ 
such that $(1-\mu)^N<\disp\frac{\gamma-\delta}{1-\delta}$ 
for all $N\geq N_{\delta}$. 
The adversary $\Adv$ repeats the following processes from Step 1 to Step 3 
at most $N_{\delta}$ times. 
\begin{description}
\item[Step 1.] The adversary 
$\Adv_2$ chooses a biometric characteristic $v\in\U$ 
uniformly at random. 
\item[Step 2.] The adversary $\Adv_2$ sends to the sampling oracle $\Samp$ and 
gets a feature element $x'$ from $\Samp$. 
\item[Step 3.] The adversary $\Adv_2$ checks whether 
$\PIC(\pi,\PIR(\alpha,x'))=\match$ or $\nonmatch$. 
\end{description}
If $\PIC(\pi,\PIR(\alpha,x'))=\match$ 
in the Step $3$ during the repetition of the above processes, 
then $\Adv_2$ finishes the processes and returns $x'$. 
\par
We say that a PT $(\pi,\alpha)$ is {\em good} if 
$\MR_{\Pi}(\pi,\alpha)>\mu$. 
If the adversary $\Adv_2$ is given a good PT $(\pi,\alpha)$, then 
the probability that $\Adv_2$ gets a feature element 
$x'$ satisfying $\PIC(\pi,\PIR(\alpha,x'))=\match$ 
during the $N_{\delta}$-time repetition of the above steps is 
greater than or equal to 
$1-(1-\mu)^{N_{\delta}}>\disp\frac{1-\gamma}{1-\delta}$. 
From $(\ref{Cheby})$, 
the probability that $\Adv_2$ is given a good PT $(\pi,\alpha)$ is 
greater than or equals to $1-\delta$. 
Therefore, we have 
\begin{align*}
\Pr[\text{$\Adv$ in $\Lam$-PAL IRR Game wins}]
>(1-\delta)\times\frac{1-\gamma}{1-\delta}=1-\gamma\;.
\end{align*}
Therefore, the result follows. 
\qed

\section{Proof of Theorem 
\ref{thm:UNARCH-UNLINK}}\label{sect:proofUNARCH-UNLINK}
In this appendix, 
we will explicitly prove Theorem \ref{thm:UNARCH-UNLINK}. 
\vskip.2cm
\noindent
{\it Proof of Theorem $\ref{thm:UNARCH-UNLINK}$.} 
It is sufficient to show that there exists an 
adversary $\Adv$ in $\{\PI,\AD\}$-UNLINK Game 
whose advantage is equal to $1-\MR_{\Pi}$. 
We define such an adversary $\Adv=(\Adv_1,\Adv_2)$ as follows. 
\begin{description}
\item[The adversary $\Adv_1$.] \quad\\
The adversary $\Adv_1$ receives $(\params,\{\PI,\AD\})$ from 
the challenger $\Ch$, 
independently chooses three biometric characteristics 
$u,u_0,u_1\in\U$ uniformly at random, 
makes queries $u,u_0,u_1$ to $\Samp$, gets 
three feature elements 
$x$, $x_0$, $x_1$ from $\Samp$, respectively, 
sends $(x,x_0,x_1)$ to $\Ch$, 
and sends a state $s'=((x,x_0,x_1),\params,\{\PI,\AD\})$ to $\Adv_2$. 
\item[The adversary $\Adv_2$.] \quad\\
The adversary $\Adv_2$ receives the state 
$s'=((x,x_0,x_1),\params,\{\PI,\AD\})$ 
and $PT=(\pi,\alpha)$ and 
$PT'=(\pi',\alpha')$ from $\Adv_1$ and $\Ch$, respectively. 
When $\PIC(\pi',\PIR(\alpha',x_1))=\nonmatch$, 
$\Adv_2$ puts $b'=0$. 
When $\PIC(\pi',\PIR(\alpha',x_0))=\nonmatch$, 
$\Adv_2$ puts $b'=1$. 
When $\PIC(\pi',\PIR(\alpha',x_0))=\match$ and 
$\PIC(\pi',\PIR(\alpha',x_1))=\match$, 
$\Adv_2$ chooses $b'$ from $\{0,1\}$ uniformly at random. 
Finally $\Adv_2$ returns $b'$. 
\end{description}
From the assumption in the statement of Theorem \ref{thm:UNARCH-UNLINK}, 
if $(\pi',\alpha')=\PIE(x_0)$ (resp. $(\pi',\alpha')=\PIE(x_1)$), then 
$\PIC(\pi',\PIR(\alpha',x_0))=\match$ 
(resp. $\PIC(\pi',\PIR(\alpha',x_1))=\match$). 
Therefore, when $b=0$, there are the following two cases in which 
$\Adv$ correctly returns $b'=0$.  
\begin{description}
\item[Case 1.] $\PIC(\pi',\PIR(\alpha',x_1))=\nonmatch$ 
\item[Case 2.] $\PIC(\pi',\PIR(\alpha',x_1))=\match$ and 
$b'=0$ is chosen from $\{0,1\}$ with probability $\disp\frac{1}{2}$. 
\end{description}
Therefore, the probability that, 
when $b=0$, 
the adversary $\Adv$ correctly returns $b'=0$ is estimated as follows: 
\begin{align*}
\Pr[\text{$\Adv$ returns $b'=0$}\;\vrule\; b=0]
=&
\mathop{\E}_{
{\footnotesize
\begin{array}{l}
(x_0,x_1)\leftarrow X(\U)\times X(\U)\\
(\pi',\alpha')\leftarrow\PIE(x_0)
\end{array}}}
\Pr\left[
\begin{array}{l}
\PIC(\pi',\PIR(\alpha',x_1))=\nonmatch
\end{array}\right]\\
&+\mathop{\E}_{
{\footnotesize
\begin{array}{l}
(x_0,x_1)\leftarrow X(\U)\times X(\U)\\
(\pi',\alpha')\leftarrow\PIE(x_0)
\end{array}}}
\Pr\left[
\begin{array}{l}
\PIC(\pi',\PIR(\alpha',x_1))=\match\\
b'=0\randomchosen\{0,1\}
\end{array}\right]\\
=&(1-\MR_{\Pi})+\frac{1}{2}\MR_{\Pi}=1-\frac{1}{2}\MR_{\Pi}\;.
\end{align*}
The success probability of $\Adv$ when $b=1$ is 
similarly estimated as follows: 
\begin{align*}
\Pr[\text{$\Adv$ returns $b'=1$}\;|\; b=1]=1-\frac{1}{2}\MR_{\Pi}\;.
\end{align*}
Hence, we have 
\begin{align*}
\ADVANTAGE^{\text{\rm $\Lam$-UNLINK}}_{\Pi,\Adv}
=\bigl|2\Pr[\text{$\Adv$ in $\Lam$-UNLINK Game wins}]-1\bigr|
=\bigl|2(1-\frac{1}{2}\MR_{\Pi})-1\bigr|
=1-\MR_{\Pi}
\end{align*}
Therefore the result follows. 
\qed

\section{Proof of Theorem \ref{thm:UNLINK-IRR}}\label{sect:proofUNLINK-IRR}
In this appendix, we will explicitly prove Theorem \ref{thm:UNLINK-IRR}. 
\vskip.2cm
\noindent
{\it Proof of Theorem $\ref{thm:UNLINK-IRR}$.} 
Put $\vep'=\disp\frac{\vep+(p_{\tau}-q_{\tau})m_{d\leq\tau}}{1-p_{\tau}}$. 
It is sufficient to show that if there exists an 
adversary $\Adv$ in $\Lam$-AL$_\tau$ IRR Game 
whose advantage is greater than or equal to $\vep'$, then 
there exists an adversary $\B$ in $\Lam$-UNLINK Game 
whose advantage is greater than or equal to $\vep$. 
Suppose that there exists an adversary 
$\Adv=(\Adv_1,\Adv_2)$ 
satisfying $\ADVANTAGE^{\text{\rm$\Lam$-AL$_\tau$ IRR}}_{\Pi,\Adv}
\geq\vep'$ in $\Lam$-AL$_\tau$ IRR Game. 
We define an adversary $\B=(\B_1,\B_2)$ 
in $\Lam$-UNLINK Game as follows. 
\par
\begin{description}
\item[The adversary $\B_1$.] \quad\\
The adversary $\B_1$ receives $(\params,\Lam)$ from 
the challenger $\Ch$, inputs $(\params,\Lam)$ into 
the adversary $\Adv_1$, and obtains a state $s$ as an output of 
$\Adv_1(\params,\Lam)$. 
Then, $\B_1$ independently chooses three biometric characteristics 
$u,u_0,u_1\in\U$ uniformly at random, 
makes queries $u,u_0,u_1$ to $\Samp$, gets 
three feature elements 
$x$, $x_0$, $x_1$ from $\Samp$, respectively, 
sends $(x,x_0,x_1)$ to $\Ch$, 
and sends a state $s'=((x,x_0,x_1),s)$ to $\B_2$. 
\item[The adversary $\B_2$.] \quad\\
The adversary $\B_2$ receives the state $s'=((x,x_0,x_1),s)$ 
and $(PT)_{\Lam}$ and $(PT')_{\Lam}$ from $\B_1$ and $\Ch$, respectively. 
When $\M_{\tau}(x_0)\cap\M_{\tau}(x_1)=\phi$, 
$\B_2$ inputs $s$ and $(PT')_\Lam$ into $\Adv_2$ and 
obtains a feature element $x'$ as an output of $\Adv_2(s,(PT')_\Lam)$. 
If $d(x_0,x')\leq\tau$, then $b'=0$, 
if $d(x_1,x')\leq\tau$, then $b'=1$, otherwise 
$b'$ is chosen from $\{0,1\}$ uniformly at random. 
When $\M_{\tau}(x_0)\cap\M_{\tau}(x_1)\neq\phi$, 
$b'$ is also chosen from $\{0,1\}$ uniformly at random. 
Finally $\B_2$ returns $b'$. 
\end{description}
When $b=0$, there are the following three cases in which 
the adversary $\B$ correctly returns $b'=0$. 
\begin{description}
\item[Case 1.] $\M_{\tau}(x_0)\cap\M_{\tau}(x_1)=\phi$ and 
$\Adv_2$ guesses a feature element $x'$ satisfying $d(x_0,x')\leq\tau$. 
\item[Case 2.] $\M_{\tau}(x_0)\cap\M_{\tau}(x_1)=\phi$, 
$\Adv_2$ guesses a feature element $x'$ 
satisfying $d(x_0,x')>\tau$ and $d(x_1,x')>\tau$, and 
$b'=0$ is chosen from $\{0,1\}$ with probability $\disp\frac{1}{2}$. 
\item[Case 3.] $\M_{\tau}(x_0)\cap\M_{\tau}(x_1)\neq\phi$ and 
$b'=0$ is chosen from $\{0,1\}$ with probability $\disp\frac{1}{2}$. 
\end{description}
Therefore, the probability that, 
when $b=0$, 
the adversary $\B$ correctly returns $b'=0$ is expanded as follows: 
\begin{align*}
&\Pr[\text{$\B$ returns $b'=0$}\;\vrule\; b=0]\\
=&
\mathop{\E}_{
{\footnotesize
\begin{array}{l}
(x_0,x_1)\leftarrow X(\U)\times X(\U)\\
PT'\leftarrow\PIE(x_0)
\end{array}}}
\Pr\left[
\begin{array}{l}
\M_{\tau}(x_0)\cap\M_{\tau}(x_1)=\phi\\
\Adv_2((PT')_\Lam)=x', d(x_0,x')\leq\tau
\end{array}\right]\\
&+\mathop{\E}_{
{\footnotesize
\begin{array}{l}
(x_0,x_1)\leftarrow X(\U)\times X(\U)\\
PT'\leftarrow\PIE(x_0)
\end{array}}}
\Pr\left[
\begin{array}{l}
\M_{\tau}(x_0)\cap\M_{\tau}(x_1)=\phi\\
\Adv_2((PT')_\Lam)=x', d(x_0,x')>\tau,\;d(x_1,x')>\tau\\
b'=0\randomchosen\{0,1\}
\end{array}\right]\\
&+\mathop{\E}_{
{\footnotesize
\begin{array}{l}
(x_0,x_1)\leftarrow X(\U)\times X(\U)\\
PT'\leftarrow\PIE(x_0)
\end{array}}}
\Pr\left[
\begin{array}{l}
\M_{\tau}(x_0)\cap\M_{\tau}(x_1)\neq\phi\\
b'=0\randomchosen\{0,1\}
\end{array}\right]\;.
\end{align*}
Since $\Pr[E_1\cap(\neg E_2\cap\neg E_3)]
\geq\Pr[E_1]-(\Pr[E_1\cap E_2]+\Pr[E_1\cap E_3])$ 
for any events $E_1$, $E_2$, and $E_3$, 
the second term is estimated as follows:
\begin{align*}
&\Pr\left[
\begin{array}{l}
\M_{\tau}(x_0)\cap\M_{\tau}(x_1)=\phi\\
\Adv_2((PT')_\Lam)=x', d(x_0,x')>\tau,\;d(x_1,x')>\tau\\
b'=0\randomchosen\{0,1\}
\end{array}\right]\\
\geq&
\disp\frac{1}{2}\left(
\Pr[
\M_{\tau}(x_0)\cap\M_{\tau}(x_1)=\phi
]-
\Pr\left[
\begin{array}{l}
\M_{\tau}(x_0)\cap\M_{\tau}(x_1)=\phi\\
\Adv_2((PT')_\Lam)=x', d(x_0,x')\leq\tau
\end{array}\right]\right.\\
&\hspace{6cm}\left.
-\Pr\left[
\begin{array}{l}
\M_{\tau}(x_0)\cap\M_{\tau}(x_1)=\phi\\
\Adv_2((PT')_\Lam)=x', d(x_1,x')\leq\tau
\end{array}\right]\right)\;.
\end{align*}
Therefore, we have 
\begin{align*}
&\Pr[\text{$\B$ returns $b'=0$}\;\vrule\; b=0]\\
\geq&
\frac{1}{2}\left(
\mathop{\E}_{
{\footnotesize
\begin{array}{l}
(x_0,x_1)\leftarrow X(\U)\times X(\U)\\
PT'\leftarrow\PIE(x_0)
\end{array}}}
\Pr\left[
\begin{array}{l}
\M_{\tau}(x_0)\cap\M_{\tau}(x_1)=\phi\\
\Adv_2((PT')_\Lam)=x', d(x_0,x')\leq\tau
\end{array}\right]\right.\\
&\hskip2cm\left.-\mathop{\E}_{
{\footnotesize
\begin{array}{l}
(x_0,x_1)\leftarrow X(\U)\times X(\U)\\
PT'\leftarrow\PIE(x_0)
\end{array}}}
\Pr\left[
\begin{array}{l}
\M_{\tau}(x_0)\cap\M_{\tau}(x_1)=\phi\\
\Adv_2((PT')_\Lam)=x', d(x_1,x')\leq\tau
\end{array}\right]+1\right)\;.
\end{align*}
By the definitions of $p_{\tau}$ and $q_{\tau}$, we have
\begin{align*}
&\Pr[\text{$\B$ returns $b'=0$}\;\vrule\; b=0]\\
\geq&
\frac{1}{2}\left((1-p_{\tau})
\mathop{\E}_{
{\footnotesize
\begin{array}{l}
x_0\leftarrow X(\U)\\
PT'\leftarrow\PIE(x_0)
\end{array}}}
\Pr[\Adv_2((PT')_\Lam)=x', d(x_0,x')\leq\tau]\right.\\
&\hskip2cm\left.
-(1-q_{\tau})
\mathop{\E}_{
{\footnotesize
\begin{array}{l}
x_1\leftarrow X(\U)\\
PT'\leftarrow\PIE(x_0)
\end{array}}}
\Pr[\Adv_2((PT')_\Lam)=x', d(x_1,x')\leq\tau]+1\right)\;.
\end{align*}
Since $\Adv_2$ is only given independent information from $x_1\in\M$, 
the probability that $\Adv_2$ guess a feature element $x'$ contained in 
the $\tau$-neighborhood to $x_1$ is at most $m_{d\leq\tau}$. 
Consequently, we have
\begin{align*}
\Pr[\text{$\B$ returns $b'=0$}\;|\; b=0]
\geq
\frac{1}{2}\Bigl((1-p_{\tau})
\Pr[\text{$\Adv$ in $\Lam$-AL$_{\tau}$ IRR Game wins}]
-(1-q_{\tau})m_{d\leq\tau}+1\Bigr)\;.
\end{align*}
We can similarly estimate the success probability of $\B$ when $b=1$ 
as follows: 
\begin{align*}
\Pr[\text{$\B$ returns $b'=1$}\;|\; b=1]
\geq
\frac{1}{2}\Bigl((1-p_{\tau})
\Pr[\text{$\Adv$ in $\Lam$-AL$_\tau$ IRR Game wins}]
-(1-q_{\tau})m_{d\leq\tau}+1\Bigr)\;.
\end{align*}
Finally, the advantage of the adversary $\B$ is calculated as follows:
\begin{align*}
\ADVANTAGE^{\text{\rm $\Lam$-UNLINK}}_{\Pi,\B}
=&\bigl|2\Pr[\text{$\B$ in $\Lam$-UNLINK Game wins}]-1\bigr|\\
\geq&\bigl|(1-p_{\tau})
\Pr[\text{$\Adv$ in $\Lam$-AL$_\tau$ IRR Game wins}]
-(1-q_{\tau})m_{d\leq\tau}\bigr|\\
=&\bigl|(1-p_{\tau})
\ADVANTAGE^{\text{\rm$\Lam$-AL$_\tau$ IRR}}_{\Pi,\Adv}-
(p_{\tau}-q_{\tau})m_{d\leq\tau}\bigr|\\
\geq&\bigl|(1-p_{\tau})\vep'-
(p_{\tau}-q_{\tau})m_{d\leq\tau}\bigr|=\vep\;.
\end{align*}
Therefore the result follows. \qed

\end{document}